
\input harvmac

\Title{RU-94XX}{$\hat A$-Genus and the Sigma Model}

\centerline{Kelly Jay Davis}
\bigskip
\centerline{Rutgers Physics}
\centerline{Rutgers University}
\centerline{Piscataway, NJ}

\vskip .3in
This set of lecture notes presents a pedantic  derivation
of  the  connection  between  the  $ {\hat A} $-genus  of
spacetime's  loop  space  and  the  genus  one  partition
function of the $ N=1/2 $ sigma model.  It concludes with
some  remarks on possible  generalizations of the $ {\hat
A} $-genus  which  follow  naturally  from  the `stringy'
point-of-view but have yet to be explored mathematically.
This  set  of   lecture   notes  is   geared   towards  a
mathematical audience unfamiliar with the $ N=1/2 $ sigma
model.

\Date{5/94}

\newsec{Introduction}

    In these  lectures we will  elucidate the  connection
between the $\hat A$-genus and a  particular sigma model.
Our plan of attack  will be to  first focus  attention on
the so-called standard  $ N=1 $ supersymmetric non-linear
sigma model.  We will then look at the standard $N = 1/2$
supersymmetric non-linear sigma model. With this  $N=1/2$
sigma model we will compute its supercurrent and then its
supercharge. Finally,  we will employ the  supercharge to
compute the genus one partition function.

    After   focusing   on   the   physics  side  of  this
discussion,  we then focus on the  mathematical side.  We
will look at the loop space  of spacetime and its tangent
bundle, then we will employ a metric over our spacetime $
M$ to create a metric over ${\cal L}M$ and the associated
Clifford algebra bundle over $ {\cal L}M $.  Finally,  we
will write down the Dirac operator over $ {\cal L}M $ and
show that its index is the  genus one  partition function
we computed above.

\newsec{Standard N=1 Supersymmetric Non-Linear Sigma Model}

    In this section of the lecture, we will introduce the
standard  $N=1$  supersymmetric  non-linear   sigma model
\ref \EWittenI{ Edward Witten,  ``Mirror   Manifolds  and
Topological Field Theory," hep-th   preprint   9112056.}.
Very generally,   the  standard  $ N = 1 $ supersymmetric
non-linear sigma  model is a  two-dimensional  `physical'
field   theory   which   describes   the  movement  of  a
one-dimensional surface in spacetime.

    More  specifically,  let  $\Sigma_{g}$  be a compact,
oriented Riemann surface, of genus  $g$,  which is also a
Riemannian manifold with metric $h$. In addition, we will
assume  that  $\Sigma_{g}$  is smooth.  Also,  consider a
smooth, `spin,' Riemannian manifold $M$ with metric  $g$,
which is also normal.  The  Riemannian  manifold $ M $ is
called the target space or  spacetime, and the Riemannian
manifold $\Sigma_{g}$ is called the world-sheet.

    Now, consider the Riemann surface $\Sigma_{g}$. As it
is,   by  hypothesis,  equipped  with  a  metric  and  an
orientation, we have the metric $h \in Met( \Sigma_{g} )$
and an orienting top-form $\epsilon \in\Gamma ( \Lambda^{
2}_{+}(\Sigma_{g})) $ associated with the metric $h$.  We
may now take a composition of  $ \epsilon $ and  $ h $ to
define a complex structure on $\Sigma_{g}$,

\eqn
\ComplexStructure
{
 \epsilon ( h^{-1}  , \cdot ) : T\Sigma_{g} \rightarrow T
 \Sigma_{g}.
}

\noindent  We  will,  from  now  on,  denote   the  above
combination as $J$ to agree with convention.   Now, as is
conventional, we may decompose $T^{*}\Sigma_{g}$ and also
$T \Sigma_{g}$ into $\pm i$ eigenspaces of $J$.  We shall
denote the $+i$ eigenspaces as $T^{1 0 *} \Sigma_{g}$ and
$T^{1 0} \Sigma_{g}$ respectively.  We  denote  the  $-i$
eigenspaces as  $ T^{0 1 *} \Sigma_{g} $  and  $  T^{0 1}
\Sigma_{g}  $ respectively.

    Now, again, if we consider $\Sigma_{g}$, then we know
that it is  `spin.'  This follows  from  the fact that if
$\Sigma_{g}$  is a  Riemann  surface which is orientable,
then all of its Stifel-Whitney classes are `trivial,'

\eqn
\StifelWhitneyClasses
{
 \eqalign
 {
  &w_{1} (\Sigma_{g}) = 0\cr
  &w_{2} (\Sigma_{g}) = 0\cr
  &w_{3} (\Sigma_{g}) = 0\cr
  &.\cr
  &.\cr
 }
}

\noindent   Therefore,  $ \Sigma_{ g } $  admits  a  spin
structure, ie.  we have a $Spin(2)$ principle bundle over
$\Sigma_{g}$ which is a lift of the  $ SO(2) $  principle
bundle over $\Sigma_{g}$ given by the metric  $h$ and the
orientation  $ \epsilon $  on  $ \Sigma_{ g } $.  So,  in
particular, this allows us  to  define  two  new  bundles
over $\Sigma_{g}$. We have $(T^{1 0 *} \Sigma_{g})^{1/2}$
and $(T^{0 1 *}\Sigma_{g})^{1/2}$ both over $\Sigma_{g}$.
These  are  the  two  spin  bundles  over  $ \Sigma_{g} $
`associated' to  $T^{1 0 *} \Sigma_{g}$  and  $ T^{0 1 *}
\Sigma_{g} $ respectively.  Also,  in the future, we will
have need of $TM$ and its complexification $T_{\bf C} M$.
So, finally, given a map $\Phi \in Map( \Sigma_{g} , M )$
we are in the situation of \fig\flabel{Diagram of Bundles
}.

%
%

    Now,  as our next task,  we  will  define  the  Fermi
fields of the $N=1$ Sigma Model. They are sections of $ (
T^{1 0 *}\Sigma_{g})^{1/2} \otimes \Phi^{*} (T_{\bf C}M)$
or of $(T^{0 1 *}\Sigma_{g})^{1/2}\otimes\Phi^{*}( T_{\bf
C}M)$. In particular we denote them as $\Psi_{+}\in\Gamma
((T^{1 0 *}\Sigma_{g})^{1/2}\otimes\Phi^{*}(T_{\bf C}M))$
and  $\Psi_{-} \in \Gamma ( ( T^{0 1 *} \Sigma_{g})^{1/2}
\otimes \Phi^{ * } ( T_{\bf C} M ) )$. Now, we define the
so-called Bose field of the model.  It is  an  element of
$Map(\Sigma_{g}, M)$. In particular we denote it as $\Phi
\in Map(\Sigma_{g}, M)$

    As, by hypothesis, $M$ is a normal manifold,  it  is,
by  definition, covered  by  open  sets $\{U\}$ such that
the map $exp_p: T_p U \rightarrow U$ is a diffeomorphisim
on each open set $U$. So, if we consider  the  Bose field
$\Phi\in  Map(\Sigma_{g},M)$, then we may compose it with
the inverse of $exp$.  This provides a map from $ \Sigma_
{g}$ to `TM.' (  Actually,  this  composition  will  only
provide a local version of this map, ie. from $U_{\Sigma_
{g}}\subset\Sigma_{g}$, where $U_{\Sigma_{g}}$ is an open
in $\Sigma_{g}$, to $T_{p}U_{M}$, the tangent space at $p
\in U_{M}\subset M$, where $\Phi (U_{\Sigma_{g}}) \subset
U_{M}$ and $U_{M}$ is an open in $M$.) So, we will denote
this composition of $\Phi$ with $exp^{-1}$ as $\phi$.

    Before we introduce the action of this $ N=1 $  sigma
model, we must  first explore  two  twisted  differential
operators over $ \Sigma_{g} $.  If we consider $\partial$
and  $ { \bar \partial } $,   the  conventional Dolbeault
operators, then we see that they have a natural action on
$(T^{1 0 *}\Sigma_{g})^{1/2}$ and $(T^{0 1 *}\Sigma_{g})^
{1/2}$. However, $\Psi_{\pm}$ are `twisted' sections of $
(T^{1 0 *}\Sigma_{g})^{1/2}$ and  $(T^{0 1 *}\Sigma_{g})^
{1/2}$ respectively. Therefore,  if we wish to extend the
action of $ \partial $ and $ {\bar \partial} $ to $ \Psi_
\pm$, then we must twist $\partial$ and  ${\bar\partial}$
by a connection on $\Phi^{*}(T_{\bf C}M)$. The connection
which we shall  employ  is  the  pull-back of the  metric
compatible, torsionless connection on $ T_{\bf C} M $. We
will denote the twisted versions of $\partial$ and ${\bar
\partial}$ as $D$ and ${\bar D}$ respectively.

    So, finally we will introduce the action of the $N=1$
sigma   model.   The   action,   from   a    mathematical
point-of-view,  may be thought of as  an energy function,
like that which arises in Donaldson theory,  or it may be
thought of as a Morse function, like that which arises in
Floer homology. In fact, the energy function of Donaldson
theory is the  action  of a  physical theory,  Yang-Mills
theory,  and the Morse function of  Floer homology is the
action of the $A$-topological sigma model.  So, with this
in mind   we introduce the action,   a map from  $ Met(M)
\times Map(\Sigma_{g},M)\times\Gamma((T^{1 0 *}\Sigma_{g}
)^{1/2}\otimes\Phi^{*}(T_{\bf C}M)) \times \Gamma((T^{0 1
*}\Sigma_{g})^{1/2} \otimes \Phi^{*} (T_{\bf C}M)) \times
\Gamma(\Lambda^{2}_{+}(\Sigma_{g}))\times Met(\Sigma_{g})
$ to $ {\bf R} $.  Given $ ( g, \Phi, \Psi_{+}, \Psi_{-},
\epsilon, h)$ the map $ S_{1} $ has value,

\eqn
\NOSNLSM
{
 \eqalign
 {
  &{S_{1}  := \int_{\Sigma_{g}}-it\bigg[  g (\Phi(z,{\bar
  z}))(\partial\phi,{\bar \partial} \phi) + ig ( \Phi (z,
  {\bar z}))  (\Psi_{-}, D\Psi_{-}) } \cr
  &{\qquad\qquad\qquad\qquad \qquad +ig(\Phi(z,{\bar z}))
  (\Psi_{+}, {\bar D} \Psi_{+}) +  {1 \over 2}R(\Psi_{+},
  \Psi_{+}, \Psi_{-}, \Psi_{-}) \bigg],} \cr
 }
}

\noindent where we have introduced a constant $t \in {\bf
R^{+} } $ and also $ R( \cdot, \cdot, \cdot, \cdot )$ the
pull-back, via $\Phi$,  of the curvature  of  the  metric
compatible, torsionless connection on $TM$.

    Now,  our next task is to  examine a special class of
symmetries  of this action.  In other  words,  we want to
examine particular $ V \in \Gamma ( T(Met(M) \times Map (
\Sigma_{g},M) \times \Gamma ((T^{1 0 *} \Sigma_{g})^{1/2}
\otimes \Phi^{*} (T_{\bf C} M)) \times \Gamma ((T^{0 1 *}
\Sigma_{g})^{1/2} \otimes \Phi^{*} (T_{\bf C} M )) \times
\Gamma (\Lambda^{2}_{+}(\Sigma_{g}))\times Met(\Sigma_{g}
)))$ such that ${\cal L}_{V} S_{1} = 0$  classically, ie.
the  Lie derivative of $S_1$ with respect to $V$ vanishes
classically.  Such a $ V $ is a so-called symmetry of the
action $S_{1}$. We will examine such $V$ by examining two
particular sections of the above bundle, $V_{+}$  and $V_
{-} $,  and their action on the fields of our theory.  We
choose $V_\pm$ such that they have the following  action,
which is parameterized by $\epsilon_{\pm}$,

\eqn
\NOSupersymmetries
{
 \eqalign
 {
  &{\cal L}_{V_{+}} \phi := i \epsilon_{+} \Psi_{-}  \cr
  &{\cal L}_{V_{+}} \Psi_{+} := -i \epsilon_{+} \Psi_{-}
  \Gamma \Psi_{+} \cr
  &{\cal L}_{ V_{+} } \Psi_{-} : = - \epsilon_{+} { \bar
  \partial} \phi \cr
  &\cr
  &{\cal L}_{V_{-}} \phi := i \epsilon_{-} \Psi_{+}  \cr
  &{\cal L}_{ V_{-} } \Psi_{+} := -\epsilon_{-} \partial
  \phi \cr
  &{\cal L}_{V_{-}} \Psi_{-} := -i \epsilon_{-} \Psi_{+}
  \Gamma \Psi_{-}, \cr
 }
}

\noindent  where  we  have  introduced  $ \Gamma $,  the
pull-back  of   the   metric   compatible,   torsionless
connection one-form on $TM$,  and $ \epsilon_{ + } $, an
arbitrary  anti-holomorphic  section of  $ ( T^{ 0 1 * }
\Sigma_{g})^{-1/2}$,  and $\epsilon_{-}$,  an  arbitrary
holomorphic section of $(T^{1 0 *}\Sigma_{g})^{ -1/2 }$.
Also, all other Lie derivatives which we did  not  write
explicitly  are  assumed  to vanish identically.  If one
explicitly checks, one finds that ${\cal L}_{V_{\pm}} S_
{1}= 0$ classically. So, $ V_{ \pm } $ are symmetries of
the action $ S_{1} $ defined above.  Furthermore,  these
are the so-called $N = 1$  supersymmetries of the action
$S_{1}$. These `simple' transformations,  in accord with
the jump from a point-particle to a  closed string,  are
responsible  for much of the  `success'  of  superstring
theory.

\newsec{ Standard N=1/2  Supersymmetric Non-Linear Sigma
         Model }

    In this  section we will  examine a variation on the
$N=1$  sigma model of last section.  We will examine the
standard $ N=1/2 $ supersymmetric non-linear sigma model
which is often called the  heterotic  sigma  model.  The
essential difference between the $ N=1 $ sigma model and
the $ N=1/2 $ sigma model is that in the $ N=1/2 $ sigma
model one simply drops $\Psi_{-}$.

    So, if we consider now the fields of the $ N = 1/2 $
sigma model,  then we  have $\Phi$ and $\Psi_{+}$  as we
simply drop $\Psi_{-}$ in this model. So,  in this model
we have an action which is simply given by dropping  the
field $\Psi_{-}$ from the $N=1$ sigma model action.  So,
the action of the $N=1/2$ sigma model is a map $S_{1/2}$
from $Met(M)\times Map(\Sigma_{g},M) \times \Gamma ((T^{
1 0 *} \Sigma_{g})^{1/2} \otimes \Phi^{*}( T_{\bf C}M) )
\times \Gamma (\Lambda^{2}_{+}( \Sigma_{g} )) \times Met
(\Sigma_{g})$ to ${\bf R}$. Given $ ( g, \Phi, \Psi_{+},
\epsilon, h )$ the map $S_{1/2}$ has value,
\eqn
\NOHSNLSM
{
 S_{1/2}:= \int_{\Sigma_{g}}-it\bigg[g(\Phi(z,{\bar z}))(
 \partial \phi, {\bar \partial} \phi) + ig (\Phi(z, {\bar
 z })) (\Psi_{+},{\bar D} \Psi_{+})\bigg].
}

    Now, our next task is to  examine a special class of
symmetries of the action $S_{1/2}$.  Again, we will want
to examine $V \in \Gamma( T ( Met(M) \times Map( \Sigma_
{g},M)\times \Gamma ((T^{1 0 *}\Sigma_{g})^{1/2} \otimes
\Phi^{*} ( T_{ \bf C } M ) )\times\Gamma(\Lambda^{2}_{+}
(\Sigma_{g})) \times Met(\Sigma_{g})))$ such that ${\cal
L}_{V}S_{1/2} = 0$ classically. We will do this, in this
case, by examining one section $V_{-}$ and its action on
the fields of our theory.  We have,

\eqn
\NOHSupersymmetries
{
 \eqalign
 {
  &{\cal L}_{V_{-}} \phi := i \epsilon_{-} \Psi_{+}  \cr
  &{\cal L}_{ V_{-} } \Psi_{+} := -\epsilon_{-} \partial
  \phi, \cr
 }
}

\noindent where, again, all Lie  derivatives which we do
not write  explicitly are assumed to vanish.  Again,  if
one explicitly checks, one finds that ${ \cal L}_{V_{-}}
S_{1/2} = 0$ classically.  So,  $V_{-}$ is a symmetry of
of the action $ S_{ 1/2 } $ defined above.  This is  the
so-called $N=1/2$ supersymmetry of the action $S_{1/2}$.

\newsec{N\"other's Theorem}

    Now,  before  going  forward,  we must  go back  and
review  a  result  of  classical  mechanics,  N\"other's
Theorem.  This  section  will  be  dedicated  to a quick
review of  N\"other's Theorem and its application to the
construction of conserved quantities. We will employ the
results of this  section,  in  accord with those of last
section,  to  derive  an  expression  for  the so-called
supersymmetric current and  supersymmetric charge of the
$N=1/2$ sigma model.

    First,     consider  $ X $   a   smooth,   oriented,
$n$-dimensional  manifold with Riemann metric $g$.  Over
$X$ assume there exists a vector bundle $E$,  our fields
will be  sections of $E$. In particular,  we will denote
an arbitrary section of $E$ as $\phi^{\alpha} \in \Gamma
(E)$.  Also, we will make the  assumption  that $ X $ is
without boundary. For our field  $\phi^{\alpha}\in\Gamma
(E)$ let us assume that we are given an action,

\eqn
\DummyAction
{
 \eqalign
 {
  &S :  \Gamma(E) \times \Gamma ( \Lambda^{n}_{+}( X ) )
  \rightarrow {\bf R} \cr
  &\qquad  ( \phi^{\alpha}, d\mu ) \mapsto\int_{X}
  {\cal L} (\phi^{\alpha}) d\mu, \cr
 }
}

\noindent where  $ {\cal L} ( \phi^{\alpha} ) \in \Gamma
(\Lambda^{0}(X))$  and $\Gamma(\Lambda^{n}_{+}( X ))$ is
the set of orienting top-forms on $ X $.  Now,  it is in
this  `generic'  situation  which  we  will  define  the
notions of  conserved current and conserved  charge  and
also derive N\"other's Theorem.

    Now, to start we must assume that the action $S$ has
a symmetry.  In other words, we assume that there exists
$V \in \Gamma( T(\Gamma (E) \times \Gamma ( \Lambda^{n}_
{+}(X))))$ such that $ {\cal L}_{V} S = 0 $ classically.
Such a  $ V $  is called a symmetry of the action $ S $.
Furthermore,  we  assume that  $ V $  has the  following
action, parameterized by $\epsilon$, on $(\phi^{\alpha},
d\mu) \in\Gamma (E) \times \Gamma (\Lambda^{n}_{+}(X))$,

\eqn
\DummySymmetries
{
 \eqalign
 {
  &{\cal L}_{V} \phi^{\alpha} := \epsilon V^{\alpha} \cr
  &{\cal L}_{V} d\mu := 0, \cr
 }
}

\noindent where $\epsilon \in \Gamma(\Lambda^{0}(X))$ is
a  small section, $| \epsilon | \ll 1$  for every $p \in
X $.  This is a  so-called  symmetry of $ S $.   For our
derivation we shall postulate such a  $ V \in \Gamma ( T
(\Gamma(E) \times\Gamma(\Lambda^{n}_{+}(X))))$ exists.

    Our  next  task is to  examine  how the action $ S $
changes under  the  action of $V$. In  doing so we shall
assume that $ {\cal L}  (\phi^{\alpha}) $ is of the form
${\cal L} ( \phi^{\alpha} ) := {\cal L} ( \phi^{\alpha},
\partial_{\rho} \phi^{\alpha}) $. So, if the term ${\cal
L}(\phi^{\alpha} ) $ is of the form  $ {\cal L} ( \phi^{
\alpha}, \partial_{\rho}\phi^{\alpha})$, then the change
in $S$ will be of the form,

\eqnn\DummyActionVariationO
$$\eqalignno
{
 \delta S
 &={\int_{X} \biggl[ \biggl({{\partial {\cal L} ( \phi^
 { \alpha }, \partial_{ \rho } \phi^{ \alpha } )} \over
 { \partial \phi^{\alpha}}}\biggr) \biggl( \delta \phi^
 {\alpha} \biggr) + \biggl( {{\partial {\cal L} ( \phi^
 { \alpha }, \partial_{ \xi  } \phi^{ \alpha } )} \over
 { \partial ( \partial_{\rho} \phi^{\alpha})}}  \biggr)
 \biggl(  \delta \partial_{\rho} \phi^{\alpha}  \biggr)
 \biggr] d\mu}&\DummyActionVariationO \cr
 &={\int_{X} \biggl[ \biggl({{\partial {\cal L} ( \phi^
 { \alpha }, \partial_{ \rho } \phi^{ \alpha } )} \over
 {\partial \phi^{\alpha}}} \biggr) \biggl( \delta \phi^
 { \alpha } \biggr) + \biggl({{\partial {\cal L} (\phi^
 { \alpha }, \partial_{ \xi  } \phi^{ \alpha } )} \over
 { \partial ( \partial_{\rho} \phi^{\alpha})}}  \biggr)
 \biggl(  \partial_{\rho} \delta \phi^{\alpha}  \biggr)
 \biggr] d\mu,} \cr
}$$

\noindent  where we have written $ {\cal L}_{ V } \phi^
{\alpha} $ as $ \delta \phi^{\alpha} $, ${\cal L}_{ V }
\partial_{\rho}  \phi^{\alpha}$  as  $ \delta \partial_
{\rho}\phi^{\alpha}$, and ${\cal L}_{V}S$ as $\delta S$
to agree with convention while assuming  that  $ \delta
\partial_{\rho} \phi^{\alpha} = \partial_{\rho}  \delta
\phi^{\alpha}$.

Now, we will evaluate $\delta S$ at a particular set of
$\phi^{\alpha} \in \Gamma(E) $.  We will evaluate it at
$\phi^{\alpha} \in \Gamma(E)$ which solve the classical
equations of motion. In other words, at $ \phi^{\alpha}
\in \Gamma(E)$ such that,

\eqn
\DummyEoM
{
 \biggl[ {{\partial {\cal L}  (\phi^{\alpha}, \partial_
 {\rho}\phi^{\alpha} )} \over {\partial \phi^{\alpha}}}
 \biggr]  =  \biggl[ \partial_{\rho}{{\partial {\cal L}
 (\phi^{\alpha}, \partial_{\xi} \phi^{\alpha}) }  \over
 { \partial ( \partial_{\rho} \phi^{\alpha}) }} \biggr]
}

So,  as we will only be evaluating $ \delta S $ at such
$\phi^{\alpha}$,  we  may   freely  employ   the  above
equation in our expression for $\delta S$.  Doing so we
have,

\eqnn\DummyActionVariationT
$$\eqalignno
{
 \delta S
 &={\int_{X} \biggl[ \biggl(  \partial_{\rho}{{\partial
 {\cal L}(\phi^{\alpha}, \partial_{\xi} \phi^{\alpha})}
 \over {\partial(\partial_{\rho}\phi^{\alpha})}}\biggr)
 \biggl(  \delta  \phi^{ \alpha } \biggr)   +   \biggl(
 {{\partial {\cal L}(\phi^{ \alpha }, \partial_{ \xi  }
 \phi^{ \alpha } )} \over { \partial  ( \partial_{\rho}
 \phi^{\alpha} )}} \biggr) \biggl(\partial_{\rho}\delta
 \phi^{\alpha}  \biggr) \biggr]
 d\mu}&\DummyActionVariationT \cr
 &={\int_{X}  \biggl[ \biggl(  \partial_{\rho}  \biggl(
 {{\partial {\cal L}(\phi^{\alpha}, \partial_{\xi}\phi^
 {\alpha} ) }  \over { \partial ( \partial_{\rho} \phi^
 {\alpha} ) } } \delta \phi^{ \alpha } \biggr)  \biggr)
 \biggr] d\mu,} \cr
}$$

\noindent  where in the first line we have  substituted
the  above  equation and in the  second  line  we  have
employed the chain rule. Now, by assumption, ${\cal L}_
V S = 0$ classically, as $V$ is a symmetry;  so,  as we
are free to choose $\epsilon$ with  arbitrarilly  small
compact support, we have,

\eqn
\DummyNoethersTheoremO
{
 \biggl[ \partial_{ \rho }  \biggl( {{\partial {\cal L}
 (\phi^{\alpha}, \partial_{\xi} \phi^{\alpha} ) } \over
 { \partial ( \partial_{\rho} \phi^{\alpha} ) }} \delta
 \phi^{\alpha} \biggr) \biggr] = 0.
}

This result is know as  N\"oether's Theorem.  It is the
basis for deriving almost all conserved  quantities  in
physics. However, $\delta\phi^{\alpha } := {\cal L}_{V}
\phi^{\alpha} := \epsilon V^{\alpha}$.   So, we have an
`arbitrary'  small  parameter in $\delta\phi^{\alpha}$.
We  are always  free  to  choose $ \epsilon $ to  be  a
constant section. This implies,

\eqn
\DummyNoethersTheoremT
{
 \biggl[ \partial_{ \rho } \biggl( {{ \partial {\cal L}
 (\phi^{\alpha}, \partial_{\xi} \phi^{\alpha} ) } \over
 {\partial (\partial_{\rho} \phi^{\alpha})}} V^{\alpha}
 \biggr) \biggr] = 0.
}

\noindent  So,  with this in mind we may introduce  $J^
{\rho} \in \Gamma(TX)$,  a so-called conserved current,
which is of the following form,

\eqn
\DummyConservedCurrentO
{
 J^{\rho} := \biggl[ {{\partial {\cal L}(\phi^{\alpha},
 \partial_{ \xi} \phi^{ \alpha } ) } \over { \partial(
 \partial_{\rho} \phi^{\alpha}) }} V^{\alpha}  \biggr].
}

\noindent  This  current  is  said  to  be conserved as
$\partial_{\rho} J^{\rho} = 0$, a result of N\"oether's
Theorem, our above result.

    Now, associated to this conserved current we have a
so-called  conserved charge,  if the  topology  of  $X$
allows.  We will assume that $ X $ is of the form $ X =
{ \bf R } \times S_{t} $, ie. $X$ admits a foliation of
$ n-1 $ dimensional leaves  $S_{t}$ parameterized by $t
\in {\bf R}$.  If this is the case, then we will have a
notion of a conserved charge.

    If we choose an arbitrary leaf  $ S_{t} \subset X $
with $t \in{\bf R}$ and a unit vector $T_{t} \in\Gamma(
TX)$  orthogonal  to the leaf $S_{t}$,  we can define a
conserved charge by,

\eqn
\DummyConservedCharge
{
 Q_{t} := \int_{S_{t}}g(T_{t},J^{\rho})\, d\mu_{S_{t}},
}

\noindent  where $S_{t}$ is the leaf we are integrating
over and  $d\mu_{S_{t}}$ is a measure on  $S_{t}$ given
by the measure on  $ X $, ie. $d\mu_{S_{t}}:= i_{T_{t}}
d\mu$.

    This   charge   is  conserved  in  the  sense  that
$ \partial Q_{t} / \partial t = 0 $,  ie. the charge is
neither created nor destroyed as `time'  $t \in{\bf R}$
goes forward.  For example, this is the method in which
the  conservation  of  electric charge is derived.

\newsec{Supercurrents and Supercharges}

    In this section  we will derive expressions for the
supercurrent and the supercharge of the $ N=1/2 $ sigma
model.  We  will  then  derive  an  expression  for the
`canonically'  quantitized supercharge of the $ N=1/2 $
sigma model.

\subsec{Supercurrent}

    In this subsection we will derive an expression for
the supercurrent of the $N=1/2$ sigma model. So, let us
look back to our derivation of the conserved current of
N\"other's Theorem. Given a symmetry generated by  $V$,
a section of $T(\Gamma (E) \times \Gamma ( \Lambda^{n}_
{+}(X)))$, we defined the conserved current by,

\eqn
\DummyConservedCurrentT
{
 J^{\rho} := \biggl[ {{\partial {\cal L}(\phi^{\alpha},
 \partial_{ \xi} \phi^{ \alpha } ) } \over { \partial(
 \partial_{\rho} \phi^{\alpha}) }} V^{\alpha}  \biggr].
}

\noindent So, our task is to reproduce this formula for
the $N=1/2$ supersymmetry transformation.  We may do so
by examining the results of the last section in  accord
with those of section three. We find,

\eqn
\NOHSuperCurrentO
{
 \eqalign
 {
  &J^{z}_{-}= \biggl[ \biggl(tg(\Phi(z,{\bar z}))({\bar
  \partial} \phi, \Psi_{+}) \biggr) \biggr] \cr
  &J^{{\bar z}}_{-} = 0, \cr
 }
}

\noindent where we have  `covariantized'  by taking the
partial  derivative  of  the  action's  integrand  with
respect to ${\bar D}\Psi_{+}$ not ${\bar\partial} \Psi_
{+}$.

\subsec{Supercharge}

    In this subsection we will derive an expression for
the supercharge of the above $ N=1/2 $ supercurrent. To
do so we must first assume that $\Sigma_{g}$,  at least
locally, admits a foliation of the form $ [0,1]  \times
S^{ 1 } $. Furthermore, we will consider a  new  set of
coordinates  on  $ \Sigma_{ g } $ such  that $ \sigma $
paramaterizes the $S^{1}$ and $\tau$ the $[0,1]$. Also,
we assume that locally $z = \sigma + i \tau$ and ${\bar
z} =\sigma - i\tau$, where $\sigma$ and $\tau$ are both
real coordinates.  So,  in   defining  charge  we  must
integrate the $\tau$ component of $J_{-}$ over a $S^{1}
$ leaf of the foliation. So, formally,

\eqn
\NOHSuperchargeO
{
 Q_{-} := \int_{S^{1}}\biggl[ {{J^{z}_{-} - J^{\bar z}_
 {-}} \over {2i}} \biggr] \, d \biggl({{(z + {\bar z})}
 \over {2}} \biggr).
}

\noindent  Now,  if  we  simply  substitute  the  above
expression we  obtained for the  supercurrent  into the
formal expression for $Q_{-}$, then we obtain,

\eqn
\NOHSuperchargeTw
{
 Q_{-} =  \int_{S^{1}} \biggl[ {{t} \over {2i}} \biggr]
 \biggl[ g( \Phi(z,{\bar z}))({\bar\partial}\phi, \Psi_
 {+})\biggr] d\biggl({{(z +{\bar z} )}\over{2}}\biggr).
}

\noindent Now, we are  further  able to write down $ Q_
{-}$ in terms of $ \sigma $ and $\tau$ by employing the
definitions of $\sigma$ and $\tau$ in terms of  $z$ and
${\bar z}$,

\eqn
\NOHSuperchargeTh
{
 Q_{-}  =  \int_{S^{1}} \biggl[  {{1} \over {2}}\biggr]
 \biggl[ tg(\Phi(\sigma, \tau)) \biggl({{\partial \phi}
 \over { \partial \tau }}, \Psi_{+} \biggr) - itg( \Phi
 ( \sigma, \tau ) )  \biggl( { { \partial \phi }  \over
 {\partial \sigma }}, \Psi_{+}\biggr) \biggr] d\sigma .
}

\subsec{Canonically Quantitized Supercharge}

    In this subsection we will derive an expression for
the  `canonically'   quantitized   supercharge  of  the
$N=1/2$ sigma model.  To do so we must remind ourselves
of the `canonical' quantization prescription. Basically
the prescription  consists of re-interpreting classical
quantities,  such as the charge,  energy,  momentum ...
as operators.  In our case the `canonical' quantization
prescription  dictates  that we make  the  substitution
$\partial \phi / \partial \tau  \mapsto  -i  \partial /
\partial \phi$. Also, the prescription dictates that we
require $\Psi_{+}^{ I } \Psi_{+}^{ J } + \Psi_{+}^{ J }
\Psi_{+}^{I} = -2g^{IJ}$  for a fixed $\tau$, where $I$
and $ J $ are $T M $ indicies and $g$ is pulled-back to
$ \Sigma_{ g } $.  So,  the  `canonically'  quantitized
supercharge is,

\eqn
\NOHCQSupercharge
{
 {\hat Q}_{-} :=  -\int_{S^{1}} \biggl[ {{t} \over {2}}
 \biggr] \biggl[ \biggl( i \Psi_{+} { {\partial}  \over
 {\partial \phi}}\biggr) + \biggl(ig(\Phi(\sigma,\tau))
 \biggl( { {\partial \phi}  \over  {\partial \sigma} },
 \Psi_{+} \biggr) \biggr) \biggr] d\sigma .
}

\newsec{Dirac Operator}

    In this section  we will review the construction of
a  Dirac operator  on a finite dimensional manifold and
its equivariant  generalization.  After this,  we  will
construct a Dirac  operator on the free loop space of a
finite  dimensional manifold and then  its  equivariant
generalization.

\subsec{Dirac Operator (Conventional Case)}

    In this subsection  we will review the construction
of a Dirac operator on $Y$ a smooth, Riemannian, `spin'
manifold of  finite dimension .  Let us take the  metric
on $Y$ to be given by $f$.  Now,  to  construct a Dirac
operator  we  must first  construct a  Clifford algrbra
bundle.

    To do so consider an arbitrary $p\in Y$. At $p\in Y
$ we may consider the tensor algebra  ${\cal T}_{p}Y :=
{ \bf R } \oplus T_{ p }Y  \oplus T_{p}Y \otimes T_{p}Y
\cdots $.   Now,  from $ {\cal T}_{p} Y $ we may define
the quotient algebra,

\eqn
\CliffordAlgebraFiber
{
 Cliff(T_{p}Y, f_{p})  :=  \biggl[ { {\cal T}_{p}Y }  /
 { ( V \otimes V + f_{p}(V,V) ) } \biggr].
}

\noindent   This  quotient  algebra  is  the  so-called
Clifford algebra of ${\cal T}_{p}Y$ and $f_p$. Now, one
may repeat this  construction for each  $ p \in Y $  to
obtain a Clifford algebra bundle $ Cliff (TY,f) $  over
$ Y $.  Also, one may  complexify this bundle to obtain
$Cliff_{\bf C}(TY,f)$.

    Now, if we choose a set of sections of $Cliff_{ \bf
C}(TY,f)$, $\Psi^{I}_{+}$,   which satisfy the Clifford
relations, then  we  may  define  the  Dirac  operator.
Locally, within a given coordinate patch on $Y$,  it is
written as,

\eqn
\DriacOperator
{
 {D \!\!\!\! /} := \biggl[ \Psi^{I}_{+}{{\partial}\over
 {\partial x^{I}}} \biggr],
}

\noindent  where  $ x^{ I } $  is  a  local  system  of
coordinates on $Y$ and the summation over the index $I$
is understood.

    Now,   let  us  attempt  to  make  an   equivariant
generalization of the above  Dirac operator.  First, we
assume that a group $ G $ acts on $ Y $. If we consider
just $ g := T_{id}G $, the group's Lie algebra, then it
also inherits a natural action on $ Y $. This action is
given by the flow of a set of elements in $\Gamma(TY)$.
So, in particular, if we choose ${\hat g} \in g$,  then
we have a corresponding element  $V_{\hat g}$  which is
in $\Gamma(TY)$ and acts on $Y$ via its flow. Now, with
such a $V_{\hat g}$ we may modify $D \!\!\!\! /$ to the
so-called equivariant Dirac operator,

\eqn
\EDriacOperator
{
 {D \!\!\!\! /}_{G}:= \biggl[ \Psi^{ I }_{+}{{\partial}
 \over{\partial x^{I}}}+  \Psi^{I}_{+}f_{IJ}V^{J}_{\hat
 g} \biggr],
}

\noindent  where, again, the summation over $I$ and $J$
is understood.  This is the  equivariant generalization
of the Dirac operator.

\subsec{Dirac Operator ( Loop Space Case )}

    In   this   subsection   we   will  generalize  the
construction of a Dirac operator,  and its  equivariant
generalization, to ${\cal L}Y := \{ \gamma \mid  \gamma
: S^{1} \to Y \}$   \ref   \EWittenII{  Edward  Witten,
``Global  Anomalies  in  String  Theory,"  Symposium on
Anomalies, Geometry and Topology 1985}. As a first step
in this process,  we must construct a  Clifford algebra
bundle over ${\cal L}Y$.

    As a first step in constructing a  Clifford algebra
bundle over ${\cal L}Y$, we will examine $T{\cal L} Y$.
So, choosing $\gamma \in {\cal L} Y$,  $ \gamma (S^{1})
\subset Y$ is a loop in $Y$.  If we wish to consider an
`infinitesimal' deformation of $\gamma(S^{1})$, then we
must specify an element of $ T_{q} Y $ for every $q \in
\gamma(S^{1})$. So, in other words, an element of $ T_{
\gamma}{\cal L}Y$ is given by a section of $\gamma^{*}(
TY)$ over $S^{1}$. So, $T_{\gamma}{\cal L}Y \cong\Gamma
(\gamma^{*}TY)$.

    So, as we have defined $ T_{\gamma} {\cal L}Y $ for
any $\gamma\in {\cal L}Y$, let us next look at a metric
on $T_{\gamma}{\cal L}Y$. Let us choose $ V_{1} $, $ V_
{2}  \in T_{\gamma }{\cal L} Y $,  ie. two  sections of
$ \gamma^{*}(TY)$. Now, at each $\sigma$ in $S^{1}$, we
may  pull-back $ f $ to obtain an  inner-product on the
vector space $ \gamma^{*}(TY) \mid_{\sigma} $.  So, the
`natural' generalization is to sum this result for  all
$\sigma \in S^{1}$,

\eqn
\LoopSpaceMetric
{
 F(\gamma)(V_{1},V_{2}) :=\int_{S^{1}}f(\gamma(\sigma))
 (V_{1}(\sigma),V_{2}(\sigma)) \, d\sigma.
}

\noindent This defines a metric $F$ on ${\cal L}Y$.  As
one may  easily check, the above metric  is  symmetric,
non-degenerate, ...

    So,  now that we have given a metric on ${\cal L}Y$
and we know what $ T{\cal L} Y $ is,  we may employ the
same  formalism   of the  finite  dimensional  case  to
create  a  Clifford  algebra  bundle  as  well  as  its
compexification $Cliff_{\bf C} ( T{\cal L} Y, F )$ over
${\cal L}Y$.  Again, we may choose a set of sections of
$Cliff_{ \bf C }(T{\cal L}Y, F)$,  which we call $\Psi_
{+}(\sigma)$,  that  satisfy   the   Clifford   algebra
relations,

\eqn
\LoopSpaceCliffordAlgebra
{
 \int_{S^{1}} \Psi^{I}_{+}(\sigma) \Psi^{J}_{+}(\sigma)
 + \Psi^{J}_{+}(\sigma) \Psi^{I}_{+}(\sigma)  =  \int_{
 S^{1}} -2 f^{IJ}(\sigma),
}

\noindent where $ I $ and $ J $ are $TY$ indicies. With
these we may define,  locally,  the Dirac operator on $
{\cal L}Y$,

\eqn
\LoopSpaceDriacOperator
{
 {D\!\!\!\! /}:=\int_{S^{1}}\Psi_{+}(\sigma){{\partial}
 \over{\partial \phi(\sigma)}},
}

\noindent   where  we  have  choosen  a  local  set  of
coordinates on ${\cal L}Y$ given by $\phi(\sigma)$. So,
this is the  generalization  of the  conventional Dirac
operator on $Y$ to the space ${\cal L}Y$.

    Now,   our   task   is  to  create  an  equivariant
generalization of $D\!\!\!\! /$ on ${\cal L}Y$. We will
do so for a very specific group action on  ${\cal L}Y$,
that of $ S^{1} $.  $ S^{1} $ acts on $ {\cal L} Y $ as
follows: If $\theta \in S^{1}$ and  $ \phi (\sigma) \in
{\cal L} Y$,  then the element $\theta$ acts as  $ \phi
( \sigma )  \mapsto  \phi ( \sigma + \theta ) $. So, by
differentiating,  the element of $ T_{\phi}{\cal L} Y $
which generates this $S^{1}$ action is ${{\partial\phi}
/ {\partial\sigma}}$.  Thus, if we employ the formalism
of the finite  dimensional  equivariant generalization,
then we have the equivariant loop space generalization,

\eqn
\LoopSpaceEDriacOperator
{
 {D \!\!\!\! /}_{S^{1}} :=  \int_{S^{1}}\biggl[ \biggl(
 \Psi_{+}(\sigma){{\partial}\over{\partial \phi(\sigma)
 }}\biggr) + \biggl( f(\phi(\sigma))({{ \partial \phi }
 \over {\partial \sigma}}, \Psi_{+}) \biggr) \biggr] \,
 d\sigma.
}

\noindent This is the equivariant generalization of the
Dirac operator ${D\!\!\!\! /}$ over ${\cal L}Y$ via the
`canonical' $S^{1}$ action.

    Now,  if we look at the  `canonically'  quantitized
supercharge of the  $ N=1/2 $ sigma model,  then we see
that ${\hat Q}_{-}$ is the operator $(-it / 2){D \!\!\!
\! /}_{S^{1}}$ on ${\cal L}M$.

\newsec{Partition Function and the ${\hat A}$-Genus}

    In  this  section  we  will   compute  the  ${ \hat
A}$-genus of the  $S^{1}$ equivariant generalization of
the Dirac operator over ${\cal L} M$, and  we will also
compute the genus one partition function of the $N=1/2$
sigma model.  Finally,  we will  compare the results of
both  calculations  and show that  they are equal,  and
then we will remark on  possible generalizations of the
$ { \hat A } $-genus  of  the   $ S^{1 } $  equivariant
generalization of the Dirac operator over ${\cal L}M$.

\subsec{${\hat A}$-Genus}

    In  this  subsection  we  will  compute  the ${\hat
A}$-genus of the  $S^{1}$ equivariant generalization of
the Dirac operator over ${\cal L} M$.   So, to start we
will remind ourselves of the operator  $(-1)^{F}$ which
acts on the complexified  Clifford  algebra bundle over
${\cal L}M$ as

\eqn
\ModTwoFermionOperator
{
 (-1)^{F}\Psi_{+}(\sigma) = - \Psi_{+}(\sigma)(-1)^{F}.
}

\noindent In other words, $ (-1)^{F} $ is the so called
chirality  operator  on ${\cal L }M$ \EWittenII. It has
a more familiar form given by

\eqn
\ChiralityOperator
{
 (-1)^{F} := i^{\infty} (\Psi^{1}_{+}(\sigma) \Psi^{2}_
 {+}(\sigma) \cdots ),
}

\noindent  where we have taken $ (-1)^{ F } $ to be the
Clifford  algebra product of all  complexified Clifford
algebra generators.

    Now, following  \ref \EWittenIII{   Edward   Witten,
``Topology and Supersymmetric  Non-Linear Sigma-Models,"
International   Workshop  on   Superstrings,   Composite
Structures and Cosmology,   University   of  Maryland at
College  Park  1987},   $ (-1)^{ F } $    is    of   use
because, when acting on a spinor bundle over ${ \cal L }
M$,  its $ \pm $  eigenspaces are the irreducible spinor
bundles $S_{\pm}$.  Also, as $(-1)^F$ anti-commutes with
all of the $\Psi_{+} ( \sigma )$,  it also anti-commutes
with $i{D\!\!\!\!  /}_{S^{1}} $,  ie. if we let  $(-1)^{
F}$ act on a spinor bundle over ${\cal L} M $,  then its
action anti-commutes with that of $i{D\!\!\!\! /}_{S^{1}
}$,

\eqn
\DriacChiralityAC
{
 (-1)^{F} i {D\!\!\!\! /}_{S^{1}} = -i{D\!\!\!\! /}_{S^
 {1}} (-1)^{F}.
}

\noindent Therefore, if $\Psi$, a spinor bundle section
over ${\cal L}M$, is an eigenspinor of $i{D\!\!\!\! /}_
{S^{1}}$ with eigenvalue $\lambda$, then $(-1)^{F}\Psi$
is also an eigenspinor of $ i{D\!\!\!\! /}_{S^{1}}$ but
with eigenvalue $-\lambda$.

    So, if $ \lambda $ above is not equal to $0$,  then
$ \Psi $ and $ (-1)^{F}\Psi $ are linearly independent.
Thus, we have two linearly independent spinors  $(1 \pm
(-1)^{F})\Psi$. $(1\pm (-1)^{F}) \Psi$ are $ (-1)^{F} $
eigenstates and also  $(1 \pm (-1)^{F}) \Psi$ both have
the same eigenvalue of

\eqn
\MHamiltonian
{
 H := ( i{ D\!\!\!\! /}_{S^{1}})^{\dagger} (i{ D\!\!\!\!
 /}_{S^{1}}).
}

\noindent So, the eigenvalues of $H$ all occur in pairs
if they are non-zero.  However, if they are zero,  then
they need not be paired. If we denote the eigenvalue of
$H$ by $ E $, then graphically we have the situation of
\fig\flabel{Eigenspectrum of H}.

%
%

    So,  finally we may  write  an  expression  for the
index  of  $ i{ D \!\!\!\! / }_{S^{1}} $.   It  is,  by
definition, the number of linearly independent $ \Psi $
with $(-1)^{F} = 1$ and $E = 0$, ie. $Dim(Ker(i{D\!\!\!
\! /}_{ S^{ 1 } }))$,  minus  the  number  of  linearly
independent $\Psi$ with $(-1)^{F} = -1$ and $E=0$,  ie.
$Dim (coKer(i{D\!\!\!\! /}_{S^{1}})) $. In accord  with
the above pairing of eigenspinors with $ E \ne 0 $,  we
may write  $ Index ( i{D\!\!\!\! /}_{S^{1}} ) := { \hat
A}$-genus of $i{D\!\!\!\! /}_{S^{1}}$ as,

\eqnn\EIndex
$$\eqalignno
{
 Index (i{D \!\!\!\! /}_{S^{1}})
 &={Tr \biggl[ (-1)^{F} e^{-{{T^{2}}\over{2}}H}\biggr]}
 &\EIndex \cr
 &={Tr \biggl[ (-1)^{F} \biggr] \biggr|_{E = 0}} \cr
 &={\biggl[ N^{(-1)^{F} = 1}_{E = 0}- N^{(-1)^{F}= -1}_
 {E = 0} \biggr],} \cr
}$$

\noindent where in the first step we introduced $ T \in
{\bf R}^{+} $,  in the second step we employed the fact
that the states of non-zero $E$ occur in pairs and thus
do  not  contribute,  and  in the  final step we simply
performed the trace and introduced $N^{(-1)^{F} = 1}_{E
= 0}$ as $Dim (Ker(i{D\!\!\!\! /}_{S^{1}}))$ and $N^{(-
1)^{F}= -1}_{E = 0}$ as $Dim(coKer(i{D\!\!\!\! /}_{S^{1
}}))$.

\subsec{Partition Function}

    In this section we compute the genus one  partition
function of the $ N=1/2 $ sigma model. So,  to start we
will  simply  define  the genus one partition function.
This is done by first assuming $\Sigma_{g}$ is of genus
one, ie.  $ \Sigma_{g} = \Sigma_{1} $, then introducing
the operator,

\eqn
\PHamiltonian
{
 {\tilde H} := 2 {\hat Q}_{-}^{\dagger} {\hat Q}_{-}.
}

\noindent Now, with this operator one may easily define
the genus one partition function. It is given by,

\eqn
\GenusOnePartitionFunctionO
{
 Z_{1} := Tr \biggl[ (-1)^{F} e^{-{\tilde H}} \biggr].
}

\noindent  Now,  if we  substitute  the  expression for
$ {\hat Q}_{-} $ in terms of ${D\!\!\!\! /}_{S^{1}}$ on
${\cal L}M$ into the above equation, then we find,

\eqnn\GenusOnePartitionFinctionTw
$$\eqalignno
{
 Z_{1}
 &={Tr \biggl[ (-1)^{ F } e^{ - { {t^2} \over {2} } H }
 \biggr]} &\GenusOnePartitionFinctionTw \cr
 &={Tr \biggl[ (-1)^{ F } \biggr] \biggr|_{E = 0}} \cr
 &={Index ( i{D\!\!\!\! /}_{S^{1}} ).}
}$$

    So,  finally   we  have  proven  that the genus one
partition function of the $N=1/2$  sigma model is equal
to $ Index (i{D\!\!\!\! /}_{S^{1}}) $.  However,  if we
employ rigidity, then we know $Index ( i{D\!\!\!\! /}_{
S^{1}}) = Index ( i{D\!\!\!\! /}) =  {\hat A}$-genus of
$ {\cal L} M $.  So,  via rigidity,  we  know  that the
partition  function at genus one of the $ N=1/2 $ sigma
model  is  the ${\hat A}$-genus of ${\cal L}M$.

\subsec{Generalization of the ${\hat A}$-Genus}

    Now, as we have proven that $Z_{1}$,  the genus one
partition function of the  $N=1/2$  sigma model, is the
${\hat A}$-genus  of the  loop space  of spacetime,  we
will make some  remarks on  possible generalizations of
the ${\hat A}$-genus.  However, before doing so we will
remark on a second method of computing $Z_{1}$.

    Now, for a fixed $ g \in Met(M) $ we may write $ Z_
{1}$ as a so-called path integral.  If we consider $Map
(\Sigma_{1}, M) \times \Gamma ((T^{1 0 *}  \Sigma_{1})^
{ 1/2 } \otimes \Phi^{*} ( T_{\bf C} M ) ) \times Met (
\Sigma_{1} )$, the $N=1/2$ supersymmetry transformation
has  a  natural  action on this space.  So,  given this
action  we  may form the quotient ${\cal I}_{1} :=  Map
(\Sigma_{1}, M)\times\Gamma((T^{1 0 *}\Sigma_{1})^{1/2}
\otimes \Phi^{*}(T_{\bf C} M)) \times Met(\Sigma_{1}) /
\sim$.   We may then  compute\footnote{ $ ^\dagger $  }
{One    must   also  include  auxiliary  fields in  the
calculation    to    obtain   the   correct   off-shell
supersymmetry.},

\eqn
\GenusOnePartitionFunctionTh
{
 {\tilde Z}_{1}:=\int_{{\cal I}_{1}} D\Phi D\Psi_{+} Dh
 \, e^{-S_{1/2}},
}

\noindent where $ S_{1/2} $ decends naturally to ${\cal
I}_{1}$ as $N=1/2$  supersymmetry is a  symmetry of the
action $S_{1/2}$, ie.  ${\cal L}_{V_{-}}  S_{1/2} = 0$.
One finds that, in fact, ${\tilde Z}_{1} = Z_{1}$. Now,
in  this  formulation  we  may  easily  generalize this
construction by considering an arbitrary genus  Riemann
surface. We first define the generalization of $ { \cal
I}_{1}$ as  ${\cal I}_{g} := Map(\Sigma_{g}, M)  \times
\Gamma((T^{10 *}\Sigma_{g})^{1/2}\otimes\Phi^{*}(T_{\bf
C} M)) \times Met ( \Sigma_{ g } ) /  \sim $,  then  we
may introduce the generalization of $Z_{1}$\footnote{$^
\dagger {}^\dagger$}{Again,  one must include auxiliary
fields.},

\eqn
\GenusgPartitionFunction
{
 Z_{g}:=\int_{ {\cal I}_{g} } D\Phi D\Psi_{+} Dh \, e^{
 -S_{1/2}}.
}

\noindent  This seems the `natural' generalization of $
Z_1$ and thus the equivariant  ${\hat A}$-genus of  $ {
\cal L} M $.  However,  we should note two  subtleties.
First,  this is the  `natural'   generalization of  the
equivariant index,  not   necessarily the normal index.
The fact that  $Z_{1}$ is also the normal  index relies
upon  rigidity  which  may  or  may  not  hold  in this
generalized   context.  (  However,  from  a  `stringy'
point-of-view it is conjectured to hold. ) Second,  the
$ N = 1 / 2 $  supersymmetries  relied upon the  global
existence  of  $ \epsilon_{ - } $.   However,  in  this
generalized context we should  not  require the section
$ \epsilon_{ - } $ to exist  globally \EWittenI.   With
these  two  provisos  the  $ Z_{ g } $  should  provide
mathematically   interesting   generalizations   of the
conventional $ S^{1} $ equivariant ${\hat A}$-genus  of
${\cal L}M$.

\listrefs
\listfigs
\bye